# Did life originate from low-temperature areas of the Universe?

Serge A. Krasnokutski

The biological molecules delivered to Earth on the board of meteorites and comets were called one of the possible causes of the origin of life. Therefore, our understanding of the routes of formation of biomolecules in space should shed the light on the possibility of the existence of habitable extrasolar planets. The large abundance of organic molecules is found in the space regions with the lowest temperature. Different routes of the organics formation in these areas were suggested. In this article, we demonstrate that complex organic molecules same as important biological molecules can be formed due to the reaction of C atoms with the mantels of molecular ices covering refractory dust grains present in the interstellar medium (ISM). Having four valence electrons, C atoms act as glue joining simple non-organic molecules and converting them into organic matter. The formation of many molecules is barrierless and thus can happen at low temperature. The barrierless reaction C + NH$_3$ + CO → NH$_2$CHCO attracts particular interest. The product of this reaction is an isomer of the central residue of a peptide chain and expected to be efficiently formed in the translucent molecular clouds. The polymerization of these molecules leads to the formation of proteins that according to some theories are life's first molecules. Considering a high abundance of atomic carbon in the ISM, we expect a high efficiency of the formation of a large variety of different organic molecules, and show why the amount of organic material formed by condensation of atomic carbon may be underestimated.

## Introduction

Diverse prebiotic and biomolecules are delivered to Earth on the boards of meteorites and comets. Analysis of meteoric materials delivered to Earth demonstrated the presence of amino acids, sugars, and nucleobases among other complex organic molecules (COMs) of extrasolar origin.[1-3] Recently, there were also two reports on the detection of proteins and peptide chains in the meteoritic samples [4,5]. Such a massive delivery of important biomolecules to Earth should have been of a major importance for the life genesis,[6] and thus the biological life, as we know it, could be rather common in our Universe.

The organic molecules are delivered to Earth not only by meteorites, material of which could be studied in the lab, but also by comets and micrometeorites, the composition of which is not precisely known. Small prebiotic molecules were detected in the coma of comets, which showed the presence of glycine along with methylamine and ethylamine.[7] However, larger organic molecules cannot be simply transferred into the gas-phase and therefore escape the detection. Therefore, to better understand what kind of biological material and at which time was delivered to Earth, it becomes important to clarify how they are formed in space.

Currently, the most probable route of organic molecules formation seems to be the surface reactions on the cold dust particles present in the molecular clouds or in the protoplanetary disks.[8,9] The two main routes are currently considered for their formation in the astrophysical environments. The first one involves the energetic processing, most commonly by UV photons or/and cosmic rays, of the ice mantels covering the refractory dust grains.[10,11] The second non-energetic route includes the addition of atomic species to the molecules from ice mantels.[12] The second route could potentially lead to a much higher amount of organics as the energetic processing not only creates the radicals required for the molecular growth but also leads to the destruction of the formed species.

Additionally, to solve the lifetime problem of interstellar dust, the low-temperature growth of dust was suggested to take place directly in the interstellar medium (ISM).[13] This could result in a simpler way of formation of organics due to the condensation of atomic gas on the cold surfaces of survived seed grains. In this case, the substantial part of simple molecules, like CO, $CO_2$, and $H_2O$, which are the main components of the ice mantels[14] could be formed due to the energetic processing of the organics formed by the condensation. This is in contrast to the first pathway, where the formation organics is due to the energetic processing of small molecules.

In all these cases the reactions leading to the formation of organics proceed while reactants have a contact with a third body. Currently, there is a very limited amount of fundamental information on such reactions. Therefore, the current models of the space chemistry have to rely on the results of the studies performed in the gas-phase or the results of quantum-chemical computations. At the same time, interaction with a third body significantly changes the pathway of reactions. The possibility to dissipate the reaction energy by transferring it to a third body could facilitate the associative reactions same as it could freeze the reaction in any intermediate state. Therefore, the reliable experimental information on the reactions processing when reactants have contact with a third body becomes very important. Currently, the most common way to study such astrochemistry reactions consists in the deposition of the reactants on the surface of the appropriate substrates, allowing them to react spontaneously or initiate the reactions via energetic processing of the ice.[12, 15-18] The reactions are most commonly monitored by recording the IR absorption spectra of the ice or by recoding the mass spectra of the desorbed species. This approach has a big advantage as it closely resembles the process occurring in space. However, for many reactions it is not possible to reproduce all physical parameters found in astrophysical environments and thus to get the clear understanding of the surface chemistry in the ISM. This is, in particular, true as some of these parameters are unknown. For instance, there are several competing interstellar dust models, which substantially vary on the composition of the dust grains.[19-23] Therefore, the knowledge on the exact chemical composition, the size, the shape, and the surface morphology of dust grains is not accurate. At the same time, it is getting difficult

Laboratory Astrophysics Group of the Max Planck Institute for Astronomy at the University of Jena, Helmholtzweg 3, 07743 Jena, Germany. Email: sergiy.krasnokutskiy@uni-jena.de.

to extract the fundamental information from the ice experiments, as the overall control on the amount of interacting reactants same as on the following reactions of the formed products is complicated.

That is why we decided to develop a new experimental technique where we can study the reactions taking place inside liquid helium nanodroplets. We employed the He droplet isolation technique pioneered by Scoles and Toennies.[24-26] In this method, liquid helium serves as an ideal chemical inert third body, that absorbs the reaction energy but does not provide any catalytic or inhibitor influence on the reaction.[27] In this method, we used the advantage of the extremely small size of the He droplets that allows us to incorporate only a single pair of reactants per droplet. Therefore, the fundamental information on the reactions can be obtained. To do so, we used the approach that was for the first time employed in the pioneering studies by Yuan Lee in crossed molecular beam techniques.[28, 29] The experimental measurements of the reaction energies are combined with the quantum-chemical computations of the reactions leading to unequivocal conclusions regarding the reaction pathways and the presence of energy barriers. In the practical application this means that we can predict which reactions will be active at low temperature on the surface of dust grains and which products and in which electronic state will be formed.

In this article we present the results obtained by this method and quantum-chemical calculations that indicate the formation of organic and biological molecules in space due to the condensation of atomic carbon together with other species, which are abundant in the ISM.

## The He droplet setup.

The He droplet apparatus is reported in earlier publications.[27, 30] Helium droplets are produced in expansion of the compressed helium (99.9999% purity) through a cooled 5 µm nozzle. The size of helium droplets was varied by changing the temperature of the nozzle. The produced He droplets, after passing the skimmer arrive to the main vacuum chamber, where several pick-up cells are installed.

In the pick-up cells the reactants are incorporated inside of He droplets by collisional pick up in the gas-phase. Chemical reactions can be monitored by different approaches. However, the mass spectroscopy and the calorimetry measurements are two most informative methods. The mass spectrometry is used to monitor the consumption of the reactants same as the formation of the new product molecules. However, this method requires ionization of the molecules. Therefore, we cannot be absolutely sure that all reactions took place before the ionization. This can be proven by the calorimetry method. It relays on the measurement of the size of He droplets before and after the reaction. The reactions energy is transferred to the helium droplets resulting in the evaporation of He atoms and each removing 7 K of energy. Therefore, the measured size difference provides us with the precise information on the energy released in the reaction of neutral reactants.[27] This information can be compared with the energy level diagram of the reaction obtained in quantum-chemical calculations. In most cases, this approach is sufficient to draw a conclusion on the product formed in the reaction. This method was employed to study $C + H_2$ and $C + NH_3$ chemical reactions discussed in this publication.

## Quantum-chemical calculations

The quantum-chemical calculations were employed to gain the information on molecular geometries and IR absorption spectra of the organic species formed at low temperatures. All new computations presented in this publications were performed at B3LYP/6-311+G(d,p) level of theory implemented in the GAUSSIAN16 package.[31] The reaction energies were obtained as the difference between the sum of the energies of reactants and the energy of the product molecules with vibrational zero-point energy corrections.

## The formation of biomolecules via C atoms condensation

Most of the dust models suggest a carbonaceous surface of the dust grains present in the ISM. When such a surface stays uncoated, the incoming carbon atoms react with the surface leading to the growth of an amorphous carbon layer and very limited amount of organics may form. The situation changes when the surface temperature of the grains becomes low. The water molecules are primarily formed directly on the surface of the dust grains in the reactions of accreted O and H atoms.[32] Therefore, if both the thermal and the photo desorption of water molecules are not efficient, the carbonaceous surface of dust grains would be quickly protected by the water ice layer. The accreted carbon atoms become weakly bounded to the water molecules and are available for chemical reactions with other accreted atoms or molecules. Currently, there are only few considered surface reactions of C atoms. For example, only simplest surface reactions of C atom are included into KIDA[33, 34] database - $C + C_n \rightarrow C_{n+1}$, $C + C_nH \rightarrow C_{n+1}H$ $n = 0 – 9$; $CH_n + H \rightarrow CH_{n+1}$, $n = 0\text{-}3$; $C + C_2H_3 \rightarrow CH_2CCH$; $C + CCN \rightarrow C_3N$; $C + CCO \rightarrow C_3O$; $C + CCS \rightarrow C_3S$; $C + CN \rightarrow CCN$; $C + HS \rightarrow H + CS$; $C + N \rightarrow CN$; $C + NH \rightarrow HNC$; $C + NH_2 \rightarrow H + HNC$; $C + NO \rightarrow OCN$; $C + NS \rightarrow S + CN$; $C + O \rightarrow CO$; $C + O_2 \rightarrow O + CO$; $C + OCN \rightarrow CN + CO$; $C + OH \rightarrow H + CO$; $C + S \rightarrow CS$; $C + SO \rightarrow S + CO$; $C + H_2 \rightarrow CH_2$; $C + NH_3 \rightarrow CH_2NH$. At the same time, C atoms are shown to be extremely reactive. The absence of the reaction barrier in the reactions of carbon with most of the carbonaceous molecules was found.[35-43] The ability of C atoms to be barrierlessly incorporated into existing C-C bonds[36, 40] creates the conditions for the growth in the molecular size and the complexity of the molecular structure at low temperatures. There are several models of the surface chemistry, where carbon atoms were involved in the formation of COMs.[44 8, 45] However, in all these models, the formation of COMs is mainly controlled by the reactions of atomic hydrogen, while at the conditions when the molecular ice mantels cover the grains, most of the hydrogen is in the molecular form. In the past, the reaction of C atoms with $H_2$ molecules was considered to be unimportant due to a large activation barrier. However, our experiential studies[27, 39] confirmed the results of quantum-chemical calculations[46] that shows the zero energy barrier for this reaction, which was found to be important for the surface chemistry of the ISM.[47] Reactions of C atoms with molecular and atomic hydrogen have noticeable difference. Addition of atomic hydrogen is expected to convert C atoms to methane, while the reactions with molecular hydrogen lead to the formation of HCH molecules. The reaction $HCH + H_2$ has rather high activation energy and therefore is not active at low temperatures. At the same time, HCH radical, similar to C atoms, can join different molecules and contribute to the formation of organic and biological molecules on the surface. An example of such a reaction is the formation of glycine molecule. One of the possible barrierless pathways for this reaction is shown in Figure 1.

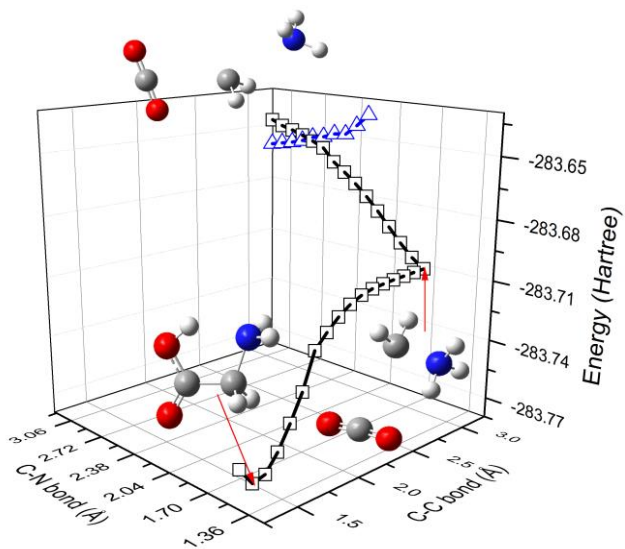

*Figure 1.* The barrierlessly pathway for the formation of glycine starting from $CO_2$, HCH, and $NH_3$ reactants found in MP2/6-311+G (d, p) calculations.[43] Squares show singlet channel while triangles are for the triplet channel of the reaction.

As can be seen from Figure 1, the singlet channel of the reactions is barrierless, while triplet channel has a small barrier. In the equilibrium, at the temperatures of solid water, the abundance of the singlet HCH is negligibly low. Therefore, the reaction should not be efficient at low temperatures. However, the surface reaction of $C + H_2 \rightarrow HCH$ releases ~321 kJ/mol of energy, which is sufficient either to help to overcome the barrier of the triplet channel of the $NH_3 + HCH + CO_2 \rightarrow$ glycine reaction or to bring the system into the singlet state with following barrierless transformation into the ground state that corresponds to the glycine molecule.

This route requires the relatively high abundance of $H_2$ molecules on the surface. Due to the low binding energy of $H_2$ and H to most of materials, their abundance on the surface is rather low even at the areas with lowest temperature. The accreted carbon atoms are expected to react immediately after landing. If carbon atoms land on the site that contain ammonia molecules, the barrierless reaction $C + NH_3 \rightarrow H_2CNH$ would take place immediately.[43] This reaction may also lead to the formation of glycine molecule. The reaction $CH_2NH + CO + H_2O \rightarrow$ glycine with energy barrier of ~152 kJ/mol, could be activated by energy 297 kJ/mol) released in the first reaction of C atoms with ammonia molecules. Moreover, the $H_2CNH$ molecule is formed in the long-living triplet state, which is about 278 kJ/mol above the ground singlet state. The energy stored in this state could help to overcome the energy barrier of the $CH_2NH + CO + H_2O$ reaction and lead to the formation of glycine. In this way, the formation of glycine is achieved due to classical chain reactions where the energy of the first reaction is used to activate the subsequent reactions. The efficiency of the chain reaction on the surface is not precisely known. There is very little experimental information on the chain surface reactions. For example, the chain reactions were identified based on the high temperatures of grains formed by the low-temperatures condensation of atoms or small molecules.[30, 48] Therefore, currently, we cannot estimate the efficiency of the molecular formation by chain reactions mechanism. Because of this completely barrierless pathways the efficiency of which would be high at any low temperature conditions became completely barrierless pathways the efficiency of which would be high at any low temperature conditions. In this respect, the formation of the central residue of the peptide chain becomes the most interesting. When $NH_3$ is located next to CO in the ice mantel, the accreted C atom reacts simultaneously with both molecules, as it follows from quantum-chemical computations. The energy level diagram of this reaction is shown in Figure 2.

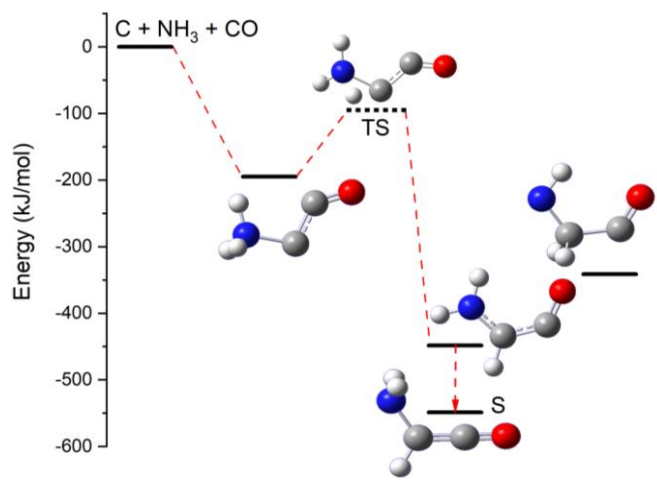

*Figure 2.* The energy level diagram of the reaction between $NH_3$, CO, and C reactants obtained at b3lyp/6-311+G (d, p) level of theory. The dashed red line shows the expected pathway of the chemical reaction. Singlet and transition states are marked by S and TS correspondingly.

It proceeds similarly to the reaction of two reactants $C + NH_3 \rightarrow HNCH_2$, which we studied experimentally.[43, 49, 50] First, the barrierless formation of the initial $NH_3CCO$ molecule takes place. The formation of this molecule is followed by the release of about 196 kJ/mol of energy. There is an energy barrier of about 100 kJ/mol to transfer the proton from nitrogen to carbon. The important point here is that this transition state is below the initial state of the reactants. Therefore, the reaction pathway does not have an energy barrier but has energy well. At low temperatures, the surface reactions may be frozen in this energy well. However, as it has been demonstrated by many different studies,[43, 49, 50] the rate of intramolecular proton transfer is much higher compared to the rate of the energy dissipation. Thus, similar to the $C + NH_3$ reaction, which we studied experimentally,[43] the intramolecular hydrogen transfer should take place well before the reaction energy is dissipated due to the interaction with the environment and reaction cannot be frozen in the well. The formation of the $NHCH_2CO$ molecule, which is the central residue of peptide chain, does not take place as it has a higher energy. This prevents the dissociation of the molecule as the singlet state of the $NHCH_2CO$ isomer is not stable. The initial triplet state of the reaction is defined by the triplet state of the C atom. The singlet states of both $NHCH_2CO$ and $NH_2CHCO$ isomers have lower energies and thus the intersystem crossing should be expected as a final step. In the singlet state, the barrierless dissociation $NHCH_2CO \rightarrow HNCH_2 + CO$ takes place, while $NHCH_2CO$ isomer is stable and its polymerization result in the formation of the glycine peptide chain.

The polymerization of $NH_2CHCO$ molecules requires the proton transfer from nitrogen to carbon, which has an energy barrier of about 250 kJ/mol. This polymerization should be simpler compared to the polymerization of glycine, which requires the abstraction of the $H_2O$ molecule and thus, rupture of several chemical bonds. Beside a much lower barrier, there are also few other advantages for the polymerization of the $NH_2CHCO$ molecule. The rate of intramolecular hydrogen transfer is extremely fast and thus it could be very efficient after the excitation of the molecule. Additionally, the importance of the tunneling for the proton transfer reactions is well known. Therefore, the polymerization $NH_2CHCO$ molecules could be quite simple. The formation of proteins could take place in ice mantels due to the photon excitation or an action of catalysts or at it can happens at later stages during the protoplanetary disk formation inside the asteroids where the liquid water could be present.

Although there is currently no clear understanding of the polymerization process, the found pathway of the $NH_2CHCO$ formation is very interesting, as it is expected to be very efficient in the astrophysical environments. When the temperature of the dust surface is below 30K that allows the CO molecules to stay on the surface, the formation of the $NH_2CHCO$ molecule should occur quite frequently. Ammonia has close to 10% of the fractional abundance in the composition of ice mantels, while for CO this value could reach up to 40%.[14] Considering also the fact that the main component of ice mantels - $H_2O$ molecules are mainly located in the inner shell, there should be many sites on the surface where both $NH_3$ and CO molecules are adjacent. Therefore, C atoms would often accrete on these sites and the following barrierless reaction $NH_3 + C + CO \rightarrow NH_2CHCO$, would account for the high efficiency of the $NH_2CHCO$ molecule formation. This process is expected to be the most productive in the translucent molecular clouds, where the temperature of the dust surface is already low enough, but most of carbon is still present in the form of atomic gas. Carbon is produced by dying stars and expelled into the ISM in the form of dust. In the ISM, this dust is atomized by the shockwaves of the supernova explosions. The gas slowly cools down and condenses. It passes through the stage of translucent molecular clouds and forms the dense molecular clouds, where new stars and planets are formed. Thus, a notable amount of $NH_2CHCO$ molecules, which is expected to be formed on the stage of translucent molecular cloud, could be delivered to planets. The delivery of these biomolecules to planets could have an impact on the origin of life.

**A hidden world of organics**

The condensation of atomic gas on the surface of dust grains could be responsible not only for a high number of biological and organic molecules but also for a high diversity of the formed molecules. Different pathways for the increasing in the complexity of the surface molecules are possible. The simplest pathway is the addition reactions when the accreted atoms are added to any species resided on the surface of the grain. For example, the hydrogenation of CO molecules with formation of methanol.[15] This is the only type of reactions possible for H atoms, as it has a single valence electron. It is also for this reason that the hydrogen cannot be added to saturated species. The situation changes considerably for the C atoms that have four valence electrons. They can be inserted into existing chemical bonds as well as added to the saturated species. Moreover, due to possibility to form four new bonds, an addition of C atoms to saturated molecules often provides molecules with the higher chemical reactivity, so that additional molecules can be added. Therefore, single C atoms can be considered as glue joining different molecules together. For instance, C atoms polymerize $C_{60}$ molecules, same as C atoms may add many other molecules to $C_{60}$.[38, 51] The same pathway was shown in the previous chapter where C atoms join the $NH_3$ and CO molecules. It is also very important that many of such reactions are barrierless and thus should be fast at low temperatures. Therefore, the accretion of C atoms on the surface of dust particles in the ISM results in the fast increase in the size and in the variety of the molecules in the ice mantels. The presence of O and N atoms during the condensation adds up to the variety. Atoms are added randomly to different molecules present on the surface. Therefore, the product of such a condensation is a complex mixture of different organic molecules and radicals. To test the material formed during such a condensation process we condense carbon atoms together with CO and $H_2O$ molecules inside liquid helium droplets and deposit the produced material on the substrates.[30] Figure 3 shows the IR absorption spectra of formed material. As can be seen from the figure, the spectrum is rather complex, containing many absorption bands in a broad spectral range. This confirms our conclusion about a wide variety of organic molecules formed by the condensation of atomic carbon. The spectrum also shows the broad distribution of the positions of the absorption bands. Here it should be noted that the spectrum was recorded at room temperature after the extraction of the substrate from the vacuum chamber. Therefore, many reactive species that are stable at low temperatures cannot be observed. Moreover, in this condensation, we used only carbon, oxygen, and hydrogen elements, which are the most abundant elements in space. However, the presence of other elements such as nitrogen, sulfur, and other less abundant elements would clearly increase the diversity of the resulting products. All these would make the resulting spectra even more complex, having larger number of absorption bands, thus reducing the intensity of each individual band and likely fusing bands to an absorption continuum. This would make it impossible to detect individual large organic molecules by spectroscopy. The common conviction is that although individual molecules cannot be detected, the presence of organics could be monitored observing absorption bands associated with specific functional groups. This is indeed the case for some simple mixtures of organic molecules. However, with an increase in sizes and number of molecules in the mixture, the position of functional groups could also be considerably altered. As can be seen in Figure 3, the bands usually associated with the CO functional group around 1700 cm$^{-1}$ are rather weak. At the same time, energy-dispersive X-ray (EDX) microanalysis of this material demonstrated a large content of oxygen in the formed material. The mass fraction of oxygen was usually higher than 25 % in all measurements.[30] Already for this material, instead of strong narrow bands, we can see the formation of the broad absorption structure 1500 – 1750 cm$^{-1}$ with few narrow bands on the top. Due to the presence of radicals and highly reactive species at low temperature, a much higher number of molecules is expected. This would result in a much larger number of absorption bands and consequently in the enhancement of the broad absorption band and the reduction of the intensity of the narrow absorption bands. To demonstrate this, we performed quantum-chemical calculations of the IR absorption spectra of the $CO_2$ molecule and few organic molecules that can be obtained from the $CO_2$ molecule as a result of addition reactions, in which the $CO_2$ fragment remains intact in the product. The obtained IR absorption spectra are shown in Figure 4. The positions of CO stretching bands are marked by the asterisk. As can be seen from this figure, the distribution of the positions of the CO stretching bands is very broad. Many radicals are stable at low temperature and have slightly different positions of absorption bands compared to the saturated molecules that dominate at higher temperatures. Additionally, at higher temperatures the variety of the molecules is reduced due to evaporation of smallest molecules. The addition of a single hydrogen atom to the OCOH molecule results in the change (shift) in frequency by about 50 cm$^{-1}$. The intermolecular interaction in the solid state, which is, in particular, important for radicals, broadens the absorption bands resulting in the overlap. The area of different functional groups vibrations also overlaps. Therefore, the organics, which is formed by condensation of atomic gas, should have very broad and weak absorption bands. Such a broad absorption feature is very difficult to detect in the observation due to the weak absolute absorption at each specific wavelength, same as due to the problems with baseline corrections. All this should finally lead to the underestimation of the amount of organics present in space. Beside the importance for the origin of life, the presence of large organics in molecular ice mantels should have a large impact on the dust formation. Currently, it is assumed that the ice mantels are quickly removed in the diffuse ISM removing any memory of the growth.[52] At the same time, it is well known that large organic molecules do not transfer into the gas phase easily. Attempts of their evaporation result in their decomposition and the formation of amorphous carbon together with few simple molecules, mainly CO, $CO_2$, and $H_2O$. Therefore, the remove of ice mantel, which is rich in organics, results in the growth of refractory grain cores.

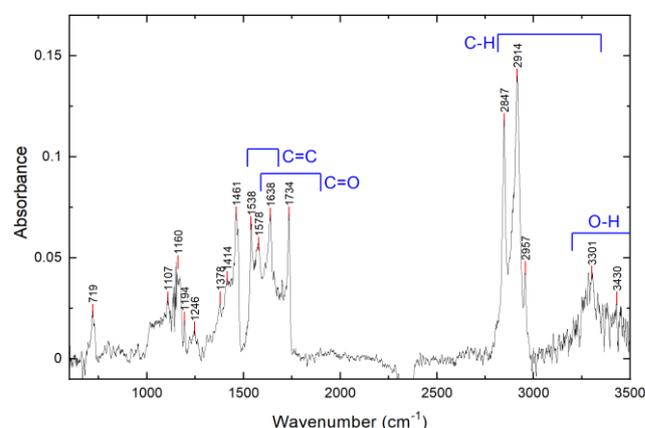

*Figure 3.* The IR absorption spectra of organic compound formed by condensation of atomic carbon together with CO, $H_2$ and $H_2O$ molecules.[30]

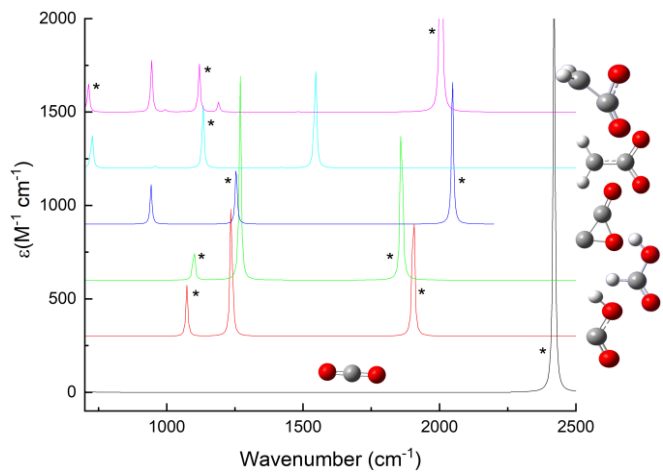

***Figure 4.*** *The infrared absorption spectra obtained in quantum-chemical calculation at b3lyp/6-311G+(d,p) level of theory. Position of CO stretching bands are shown by asterisks.*

## Conclusions

In this paper, we demonstrated a great importance of the reactions of carbon atoms accreted on the cold surfaced of interstellar dust grains for the formation of different organic and biological molecules. The carbon atoms act as a molecular glue joining different molecules together and converting inorganic to organic material. Due to an extremely high reactivity of C atoms, they are expected to react with most of the molecules found in the molecular ices covering the dust grains present in the ISM. This produces a large variety of organic molecules resulting in the overlap of their absorption bands. The resulting broad and weak absorption features of such organic mixtures became difficult to detect in the observational spectra. This could lead to the underestimation of the amount of organic material present in space.

**Acknowledgements**
The authors are grateful for the support by the Max Planck Society and the DFG (contract No. KR 3995/4-1).

**References**
[1] S. Pizzarello, and J. R. Cronin, Nature **394** (1998) 236.
[2] P. Ehrenfreund *et al.*, Geochim. Cosmochim. Acta **66** (2002) A208.
[3] N. Fray *et al.*, Nature **538** (2016) 72.
[4] L. Justin *et al.*, https://chemrxiv.org...
[5] M. W. M. a. S. D. a. J. E. M. McGeoch, (https://arxiv.org/abs/2002.11688)
[6] B. K. D. Pearce *et al.*, Proc. Natl. Acad. Sci. U.S.A. **114** (2017) 11327.
[7] K. Altwegg *et al.*, Sci. Adv. **2** (2016) e1600285.
[8] E. Herbst, and E. F. van Dishoeck, Annu. Rev. Astron. Astrophys. **47** (2009) 427.
[9] A. G. G. M. Tielens, Reviews of Modern Physics **85** (2013) 1021.
[10] M. P. Bernstein *et al.*, Nature **416** (2002) 401.
[11] G. M. M. Caro *et al.*, Nature **416** (2002) 403.
[12] H. Linnartz, S. Ioppolo, and G. Fedoseev, Int. Rev. Phys. Chem. **34** (2015) 205.
[13] B. T. Draine, Cosmic dust - near and far, ASP Conference Series, ed. Th. Henning, E. Grün, and J. Steinacker **414** (2009) 453.
[14] K. I. Oberg, Chem. Rev. **116** (2016) 9631.
[15] K.-J. Chuang *et al.*, Mon. Not. R. Astron. Soc. **467** (2017) 2552.
[16] P. D. Holtom *et al.*, Astrophys. J. **626** (2005) 940.
[17] A. K. Eckhardt *et al.*, Angew. Chem. Int. Ed. **58** (2019) 5663.
[18] A. M. Turner *et al.*, Astrophys. J. **896** (2020)
[19] A. G. Li, and B. T. Draine, Astrophys. J. **554** (2001) 778.
[20] A. G. Li, and J. M. Greenberg, Astron. Astrophys. **323** (1997) 566.
[21] J. S. Mathis, and G. Whiffen, Astrophys. J. **341** (1989) 808.
[22] D. B. Vaidya, R. Gupta, and T. P. Snow, Mon. Not. R. Astron. Soc. **379** (2007) 791.
[23] A. P. Jones *et al.*, Astron. Astrophys. **558** (2013)
[24] J. P. Toennies, and A. F. Vilesov, Angew. Chem. Int. Ed. **43** (2004) 2622
[25] W. Schollkopf, and J. P. Toennies, Science **266** (1994) 1345.
[26] S. Goyal, D. L. Schutt, and G. Scoles, Phys. Rev. Lett. **69** (1992) 933.
[27] T. K. Henning, and S. A. Krasnokutski, Nat. Astron. **3** (2019) 568.
[28] Y. T. Lee, Science **236** (1987) 793.
[29] Y. T. Lee, Angew. Chem. Int. Ed. Engl. **26** (1987) 939.
[30] S. A. Krasnokutski *et al.*, Astrophys. J. **847** (2017) 89.
[31] M. J. Frisch *et al.*, Wallingford, CT, 2016).
[32] E. F. van Dishoeck, E. Herbst, and D. A. Neufeld, Chem. Rev. **113** (2013) 9043.
[33] V. Wakelam *et al.*, Astrophys. J. Suppl. S. **199** (2012) 21.
[34] Wakelam et al. 2012, http://kida.obs.u-bordeaux1.fr
[35] R. I. Kaiser *et al.*, J. Chem. Phys. **110** (1999) 10330.
[36] R. I. Kaiser *et al.*, J. Chem. Phys. **110** (1999) 6091.
[37] S. A. Krasnokutski, and F. Huisken, J. Chem. Phys. **141** (2014) 214306.
[38] S. A. Krasnokutski *et al.*, J. Phys. Chem. Lett. **7** (2016) 1440.
[39] S. A. Krasnokutski *et al.*, Astrophys. J. Lett. **818** (2016) L31.
[40] S. A. Krasnokutski *et al.*, Astrophys. J. **836** (2017) 32.
[41] I. W. M. Smith, Angew. Chem. Int. Ed. **45** (2006) 2842.
[42] D. Qasim *et al.*, Rev. Sci. Instrum. **91** (2020) 054501.
[43] S. A. Krasnokutski, C. Jäger, and T. Henning, Astrophys. J. **889** (2020) 67.
[44] M. A. Requena-Torres *et al.*, Astrophys. J. **672** (2008) 352.

[45] M. Ruaud *et al.*, Mon. Not. R. Astron. Soc. **447** (2015) 4004.
[46] L. B. Harding, R. Guadagnini, and G. C. Schatz, J. Phys. Chem. **97** (1993) 5472.
[47] M. Simončič *et al.*, A&A **637** (2020) A72.
[48] T. Wakabayashi *et al.*, J. Phys. Chem. B **108** (2004) 3686.
[49] P.-T. Chou *et al.*, J. Phys. Chem. A **105** (2001) 1731.
[50] J. Lee, C. H. Kim, and T. Joo, J. Phys. Chem. A **117** (2013) 1400.
[51] S. A. Krasnokutski *et al.*, Astrophys. J. **874** (2019) 149.
[52] A. Ferrara, S. Viti, and C. Ceccarelli, Mon. Not. R. Astron. Soc. Lett. **463** (2016) L112.